Review Article

# The development and applications of ultrafast electron nanocrystallography

Chong-Yu Ruan*, Yoshie Murooka, Ramani K. Raman, Ryan A. Murdick, Richard J. Worhatch, Aric Pell

*Department of Physics and Astronomy, Michigan State University, East Lansing, Michigan 48824, USA*
*\*Email address: ruan@pa.msu.edu*

**Abstract**

We review the development of ultrafast electron nanocrystallography as a method for investigating structural dynamics for nanoscale materials and interfaces. Its sensitivity and resolution are demonstrated in the studies of surface melting of gold nanocrystals, nonequilibrium transformation of graphite into reversible diamond-like intermediates, and molecular scale charge dynamics, showing a versatility for not only determining the structures, but also the charge and energy redistribution at interfaces. A quantitative scheme for three-dimensional retrieval of atomic structures is demonstrated with few-particle (< 1000) sensitivity, establishing this nanocrystallographic method as a tool for directly visualizing dynamics within isolated nanomaterials with atomic scale spatio-temporal resolution.

# INTRODUCTION

Nanostructured materials are characterized by the emerging properties that are associated with the size reduction. The fundamental structure aspect on 1-100nm has been an intriguing subject. One well-known issue is the emergence of non-crystallographic structural types (Ino & Ogawa, 1967; Wales, 2000) with a tendency to form closed shells exhibiting magic numbers in size distribution (Eberhardt, 2002; Mackay, 1962). The morphology and lattice parameters of these nanocrystals depend highly on the supporting substrate and surface modification (Whetten et al., 1996), leading to contraction (Zanchet et al., 2000) and twin boundaries formation (Marks, 1994) resulting from the relaxation of surface strains. Concurrent with these new structural forms, novel properties emerge due to enhanced coupling between the electronic degrees of freedom and atomic structures on the nanometer scale. The capability of engineering specific properties by nano-synthesis and self-assembly offers interesting prospects of using nanoparticles as the building blocks for advanced devices with high specificity, speed, and density in diverse areas, such as electronics (Klein et al., 1996), photonics (Alivisatos et al., 1996), magnetism (Awschalom et al., 1992), catalysis (Haruta, 1997; Hakkinen et al., 2003), and sensing (Shipway et al., 2000). Understanding the underlying size and shape-dependent processes, such as phase change, transport, and chemistry, requires not only capability of imaging the materials and interfaces, but also following the redistribution of atoms and charges in real time. The development of ultrafast electron diffraction (UED) (Srinivasan et al., 2003; Dwyer et al., 2006; Dudek & Weber, 2001), crystallography (UEC) (Ruan et al., 2004), and microscopy (UEM [Lobastov et al., 2005]; DTEM [King et al., 2005]), which aim to unite the spatial resolution of diffraction with the temporal resolution of a femtosecond (fs) laser in an optical pump/diffraction probe setup, provides a powerful new way of studying the dynamical processes in materials at the most elementary level.

# ULTRAFAST ELECTRON NANOCRYSTALLOGRAPHY

## Concept

Here we present the development of the ultrafast electron nanocrystallography (UEnC), a special implementation of UEC, optimized to achieve high sensitivity and efficient data acquisition for quantitative dynamical investigation of nanoscale phenomena on the ultrafast (fs-ns) time scale. With the transformative development of time-resolved electron diffraction and the recent progress in instrumentation and experimental techniques such as the use of short flight distance (Siwick et al., 2003) and high acceleration voltage (Cao et al., 2003) to generate fs electron pulses, radio frequency bunching (Anderson et al., 2005; Oudheusden et al., 2007) for high brightness, photofield emission (Hommelhoff et al., 2006) for potentially high spatial coherence, and the wave-front tilting scheme (Baum & Zewail, 2006) to reduce velocity mismatch, the prospect of time-resolved structural investigation on sub-picosecond timescale for complex nanostructured materials is on the horizon. Yet, the progress towards quantitative structural determination is so far hindered, due to lack of understanding of the systematic effects introduced into the time-resolved diffraction patterns, and the absence of a reliable scheme to invert the adequately sampled excited–state diffraction data to afford the three-dimensional (3D) atomic structural determination. UEnC development aims to address these issues, with the development of new schemes for data acquisition and reduction along with a robust 3D refinement. Here, we introduce the probing concept of UEnC based on nanometer scale "powder-diffraction", where the nanostructures are dispersed sparsely on a solid substrate as shown in Figs. 1(a) – (b). While it is possible to disperse nanostructures on ultrathin amorphous film in a transmission electron microscope grid, it is found that the instability of the film can introduce artifacts in the diffraction data, or even get damaged in the high temperature regime where the film is subjected to intense laser illumination. To avoid such vulnerability, and to generalize the study to include interfacial dynamics (see Applications section), we choose to anchor the nanoparticles on a firm solid surface and use a grazing incidence electron probe to interrogate particles. In addition, we employ a buffer ad-layer composed of a self-assembled monolayer (SAM) of aminosilane molecules to elevate the nanoparticle above the solid surface, which serves three important purposes (Ruan et al., 2007) – (a) it suppresses the scattering from the substrate that would otherwise produce a strong background signal, thus overwhelming the diffraction signal from the nanostructures, (b) allows transmission diffraction signals to be collected unobstructed over a $2\pi$ solid angle, and (c) controls the rates of heat and charge dissipation of the nanoparticles. To ensure that the outgoing diffracted beams have a relatively low probability of being intercepted by another nearby nanoparticle, an optimum particle density of $\sin^2(\theta_i)/d^2$ is desirable, where $d$ is the diameter of the nanoparticle and $\theta_i$ is the electron incidence angle. Fig 1. (c) shows the SEM image of a typical sample desirable for UEnC satisfying all the above criteria using 40nm Ag nanoparticles sparsely dispersed (~10 $\mu m^{-2}$) on a Si-111 substrate on top of the SAM layer. The diffraction signal from this sample (1 minute signal integration time) is shown in Fig. 1(d) and possesses sufficient signal-to-noise ratio (SNR) for reliable quantitative studies, despite the nanoparticles occupying less than 1% of the area. This is in contrast to X-ray and neutron diffraction approaches in which the required samples must typically have a high volume density and be μm to mm in size due to the low scattering cross-sections of X-rays and neutrons, which is ~ $10^4$-$10^5$ times smaller than that of that of electrons.

**Apparatus**

Our experimental setup shown in Fig. 2(a) consists of a fs-laser and a proximity coupled photoemission electron gun arranged in a pump-probe geometry, integrated with a ultrahigh vacuum (UHV) chamber. The 800nm fs laser pulse acts as a "pump" to initiate the reaction of interest within the sample inside the UHV chamber, which is then probed by the fs electron pulses. These electron pulses are generated via photoemission from a thermally evaporated 40nm silver film, triggered by a 266nm fs laser pulse that is obtained by splitting and frequency-tripling a part of the pump pulse. Since the initiating 266nm laser pulse for generating probe electron pulse is derived from the pump laser pulse, the jitter between the pump and the probe is controlled by photoemission process, which has been clocked at the fs level or below (Cavalieri et al. 2007). The arrival of the probing electron pulse at the sample relative to that of the pump is controlled using a translational optical delay stage, allowing us to map the temporal dynamics of the reaction. The laser system (*Spectra Physics*, Ti:Sapphire, Regenerative Amplification) is capable of delivering 45fs pulses centered at 800nm with pulse energy of 2.5mJ at a 1kHz repetition rate. The frequency doubled (400nm) and tripled (266nm) pulses derived from this can attain energies up to 250 μJ and 40 μJ respectively, and can also be used as the pump-pulse.

The proximity-coupled electron source (Fig. 2(a)) delivers 30 keV electron pulses (anode extraction field 6.5 kV/mm), which subsequently pass through apertures and a magnetic lens system, capable of short-distance focusing down to as small as 5 μm at the sample ~ 5 cm from the electron source (Fig 3). The first short-distance fs photoelectron gun was designed by Miller's group (Siwick et al., 2003) to achieve high electron densities (up to $10^4$ e⁻/packet) for single-shot diffraction. For kHz multi-shot UEnC experiments however, electron counts of $10^3$ e⁻/packet are generally sufficient. Our efforts have been directed towards focusing the electron pulse down to the 1 μm level, but maintaining the fs time structure, in order to reduce the group velocity mismatch (GVM) (Williamson & Zewail, 1993), which would eventually allow time-resolved study of a single isolated nanoparticle. To demonstrate the temporal resolution of this setup, we examine the response from a photoexcited highly ordered pyrolytic graphite (HOPG) sample. In HOPG, the $E_{2g}$ optical phonon mode, which corresponds to an inter-layer shearing motion between the graphitic layers, is known to be strongly coupled to fs laser excitation (Mishina et al., 2000; Ishioka et al., 2001). Following photoexcitation, we are able to observe the sub-ps generation of surface charges in HOPG and their subsequent oscillations with a period consistent with the $E_{2g}$ mode, as shown in Fig. 4. This is attributed to the charge accumulation at the outer surface of the graphite driven by photoinduced tunneling through the surface work function barrier, forming a two-dimensional electron gas while leaving the positive ions behind in the photoexcited region. The surface charge dynamics are closely coupled to the transport characteristics perpendicular to the basal plane of graphite that is modulated by the sub-surface phonons. These optically induced redistribution of charges and corresponding atomic rearrangements will be discussed in detail in the later sections.

In addition to this pump-probe setup, the UHV chamber is also equipped with a sample transfer and positioning system, and other characterizing and cleaning instruments, as shown in Fig. 2(b). To maintain the high vacuum environment, we incorporated a loadlock extension to the UHV chamber that allows us to transfer samples directly from ambient environment into the chamber. To improve the electron detection efficiency, we implement an intensified CCD camera system capable of single electron detection. For precision temperature and position control, a home-built cryo-cooling 5-axis goniometer is used to house the sample holder with temperature regulation from 20 to 300 K. An HOPG substrate is placed just above the sample for *in situ* calibration of the pump-probe alignment in both space and time. A sputtering gun, telefocus infrared lamp, and a gas doser system are also available for surface cleaning and preparation. Finally, a quadrupole mass spectrometer and a Low Energy Electron Diffraction/Auger system are available for characterizations.

**Sample preparation**

In this section, we discuss in detail the surface preparation that makes quantitative nanoparticle diffraction possible. As previously discussed, pre-requisite for obtaining quantifiable diffraction pattern relies on adequate molecular buffering to allow controlled dispersion of nanoparticles and suppression of background. The preparation of UEnC samples largely follows recipes developed for molecular electronics (Sato et al., 1997a; Shipway, 2000) to construct a nanoparticle-SAM interface. The five major steps of this sample preparation procedure are described in Fig. 5. (a) the Si substrate is functionalized with a thin hydroxylated oxide layer (1-5 nm) following a modified RCA procedure (Kern & Puotinen, 1970). (b) The functionalized Si substrate is then immersed into a solution of surfactant molecules with an amino ($-NH_2$) tail group and a polar surface group ($-Si(OMe)_3$) for 20 minutes, the SAM is formed on the Si surface through covalent 'silanization'. The $-Si(OMe)_3$ head group reacts directly with hydroxyl groups on the Si, leading to surface coverage of an amino-terminated SAM. (c) Formation of a more compact, order self-assembled monolayer is driven by lipophilic interactions between alkane moiety, which typically takes hours and can only proceed under proper temperature. This is achieved by heating the sample in an oven at 80°C filled with high purity dry nitrogen (Sato, et al., 1997a). (d) The amino-group termination is then protonated to carry positive charge by immersion in a weak acid environment, and it can then serve as the anchoring site for negatively charged colloidal Au clusters (Schmitt et al., 1999; Liu et al., 2002; Shipway et al., 2000). (e) The sample is then removed from the solution, blown dry with nitrogen and ready for characterization.

Controlled dispersion of nanoparticles is achieved by adjusting the acidity and hydrophobicity of the solution. During surface functionization of the SAM, the density of $–NH_2^+$ anchoring sites can be varied by adding controlled amount of a weak acid, such as acetic acid (Bhat et al., 2002), which in turn determines the final optimal density of nanoparticle dispersion. Fine control of particle density can also be achieved by adjusting the hydrophobicity in solution. As the amino-terminated surface is hydrophilic, adding the proper amount of ethanol can shift the overall hydrophobicity of the solution (Westcott et al., 1998). The effects from varying the acidity are shown in Fig. 6(a) and (b), where the addition of acetic acid in solution dramatically changes the density of nanoparticles coverage, from 70 $\mu m^{-2}$ at pH 7 to 300 $\mu m^{-2}$ at pH ~ 2.0. Increasing ethanol concentration reduces the hydrophobicity of the solution, which enhances the mobility of nanoparticles, resulting in different coverage of the nanoparticles on the surface. Fig. 6(c)-(f) shows the change in the nanoparticle density for four different ratios of the mixture of ethanol and deionized water, varying from $160/\mu m^2$ to $508/\mu m^2$ as the ethanol concentration increases.

**Data Reduction**

In this section, we describe the procedure for data reduction from the diffraction images obtained using this specific 'powder diffraction' geometry. This is a critical part in the development of UEnC, as any quantitative structural analysis must begin with reliable data reduction procedures to retrieve a platform-independent structure function *S(s)* or pair distribution function *G(r)*. Here, we define the momentum transfer wavevector *s* in the reciprocal space as $s = (4\pi/\lambda)\sin(\theta/2)$ with θ denoting the scattering angle

The details of data reduction procedures usually depend on the specifics of the experimental technique (electron, X-Ray, neutron), but in principle, generally involve distilling the intensity data *I(s)* into the structure function *S(s)*, by removing the incoherent background and atomic self-scattering contributions followed by proper normalization with the atomic scattering factor. *S(s)* readily yields the pair distribution function *G(r)* through Fourier analysis. An ultimate test of the validity of the procedure is judged by the reproducibility of the *G(r)* to properly describe a known atomic structure. The data reduction procedure for UEnC is outlined in Fig. 7. The first step is to obtain the total scattering intensity function *I(s)* by radial averaging of the Debye-Schererr rings in powder diffraction patterns. To do this, we need to locate the center of these rings, which is done by symmetrizing the radially averaged diffraction curves over different angular sectors of the rings. The radial averaging procedure considers the electron incidence angle to the sample surface, sample stage orientation, camera distance, and the non-vertical incidence of the direct beam into the imaging plane. Once the ring center is determined, and the I(s) is obtained, a background removal procedure is to obtain the structure-related molecular scattering function $I_M(s)$, which is then used to deduce the radial distribution function *R(r)*. The formulation that links *I(s)* to *R(r)* is given below (Warren, 1990):

$$I_S(s) = I_A(s) + I_M(s), \qquad (1)$$

$$I_A(s) = N\langle f \rangle^2 + I^{inelastic}(s), \qquad (2)$$

$$I_M(s) = N\langle f \rangle^2 \int_0^\infty R(r) \frac{\sin sr}{sr} dr. \qquad (3)$$

The incoherent background $I_A$ is composed of contributions from atomic self-scattering $N\langle f \rangle^2$ and inelastic scattering $I^{inelastic}$; $R(r) = 4\pi r^2 \rho(r)$, where $\rho(r)$ is the atomic number density. For electron scattering, both functions are largely smoothly varying, and can be described by polynomial expansions (Doyle & Turner, 1968). Here *f* is the atomic scattering factor, *N* is the total number of atoms scattered. Note that both $I_A$ and $I_M$ depend strongly on the scattering geometry. For example, in a reflective scattering experiment, both terms are modified from their canonical form in transmission geometry by an "absorptive" shape function. However, since $N\langle f \rangle^2$ also is modified by the same absorption, we can still deduce the structural function, $S(s) = I(s)/N\langle f \rangle^2$ by using $I_A$ to represent $N\langle f \rangle^2$ followed by the addition of a low-order polynominal background to account for any offset from $N\langle f \rangle^2$ and the contribution from $I^{inelastic}$. Finally, we deduce the pair distribution function, *G(r)* by performing a sine transform:

$$G(r) = \frac{2}{\pi} \int_0^\infty s[S(s)-1]\sin(sr)ds. \qquad (4)$$

The ground state $I_M(s)$ for a 2nm Au nanoparticle following the above data reduction procedure is shown in Fig. 8 and compared with the simulated curves for known cuboctahedral, decahedral, and icosahedral structures. By inspection, we find a clear resemblance between the experimental $I_M(s)$ curves and the ideal cuboctahedral one, which are determined based on finite fcc structural order. The $G(r)$ obtained from these experimental and theoretical curves is shown in Fig. 9 and compared to atom-atom distance table for an fcc lattice. We find the experimental data to be in excellent agreement with the fcc structures, to within 0.05 Å.

To test the robustness of this data reduction procedure, we employ several different background fitting routines to $I(s)$ to extract $I_A$, and compute the corresponding $G(r)$, as shown in Fig. 10(a). The difference comparison in $G(r)$ in Fig. 10(b) shows negligible difference between G(r) estimated from different background fitting routines. The only discrepancy that goes beyond the level of intrinsic noise level of finite-range Fourier analysis is at small $r$ region with weak density peaks. These peaks arise from the diffusive molecular scattering, and might not be properly treated by the background reduction procedures aimed for treating crystalline data. Nevertheless, these density peaks are generally incommensurate with fcc structures, and can thus be ignored during refinement. Thus, the exact nature of the fitted background does not affect the data reduction significantly, so long as the background function is smooth over the entire fit range. This internal consistency in the reduction of $S(s)$ and $G(r)$ is remarkable, making the subsequent quantitative structural refinement and modeling reliable.

Various naming conventions in formulating Debye-Scherrer diffraction exist, depending on the technique. The definition of $s = (4\pi/\lambda)\sin(\theta/2)$ used here is consistent with the convention in the noncrystalline electron diffraction community and is identical to $Q$, which is widely adopted by X-ray and neutron diffraction communities. The gas phase electron diffraction community prefers the use of the molecular function (Hargittai & Hargittai, 1988):

$$M(s) = \frac{1}{N\langle f \rangle^2} \sum_{i \neq j} f_i f_j \frac{\sin(s \cdot r_{ij})}{s \cdot r_{ij}} e^{-s^2 l_{ij}^2/2} \qquad (5)$$

, which is equivalent to $S(s)$ - 1 in Eqn. (4). Note, here $M(s)$ decays as $1/s$ due to spatial averaging of randomly oriented nanoparticles. It is customary to use $sM(s)$, as a normalized form, to represent the molecular diffraction component. In addition, $G(r)$ is frequently called the "modified radial distribution function" (mRDF), or $f(r)$, as $G(r) = 4\pi r\rho(r) = R(r)/r$, is a reduced form of radial distribution $R(r)$. $G(r)$ and $sM(s)$ are eloquently related through a Fourier-sine relationship:

$$G(r) = \frac{2}{\pi}\int_0^\infty sM(s)\sin(sr)ds. \qquad (6)$$

The self-reliance feature of this data reduction procedure, with no dependence on an explicit calculation of $f(s)$ in the reflection geometry, is tremendously helpful to treat the UEnC data as the electron incidence angle is *in situ* optimized according to the specific configuration of the interface. Total scattering, hence the value of $N$, is determined by fitting $I_A$ to the canonical form of $<f(s)>^2$ in transmission geometry for sufficiently large values of $s$, where the contribution from $I^{inelastic}$ and absorption effects are negligible.

To model the excited-state data, in addition to considering the structure related changes in the diffraction pattern, it is also necessary to treat the 'distortion effect' arising from Coulomb refraction of the electron beam due to a photoinduced surface potential. Since the probing electron is a charged particle, it is deflected in the presence of a transient photovoltage generated at the interface due to the photoinduced redistribution of charges, causing a collective shift in the diffraction pattern. By measuring the magnitude of this Coulomb refraction shift, UEnC can measure the transient photovoltage associated with charge transfer at surface. An inhomogeneous shift of rocking curve from reflective electron diffraction due to surface charge transfer has been previously discussed by Peng, Dudarev, and Whelan [Peng, Dudarev & Whelan 2004]. In powder diffraction, this Coulomb refraction shift is usually perpendicular to the surface, and distorts the circular symmetry of the ring pattern toward the shadow edge, causing a swing of $s = 0$ away from the direct beam. This introduces a phase shift into $sM(s)$, thus destroys the Fourier-sine relationship between $sM(s)$ and $G(r)$. The phase shift, however, is implicit when performing a complex transform to deduce G(r) based on the overall amplitude. We exploit this idea to implement an iterative "Fourier phasing" scheme, which re-symmetrizes the diffraction pattern by compensating for the Coulomb refraction shift to remove the phase shift in $sM(s)$. The magnitude of compensation necessary thus yields the effective surface potential, $V_s$. The re-symmetrization of the diffraction pattern takes into account the angular dependence of the Coulomb refraction and assumes a rather simple form in the case of dipolar Coulomb field (Fig 11):

$$\frac{V_s}{V_0} = \frac{\left(\theta_o \delta - \theta_i \theta_o + \delta^2/2\right)^2 - \theta_i^2 \left(\theta_o + \delta\right)^2}{\left(\theta_i + \theta_o\right)^2} \quad (7)$$

, where $V_s$ is essentially a fitting parameter used to produce the observed shifts $\delta = \left(\theta'_o - \theta_o\right)$ of the Bragg peaks, and $V_0$ is the incident energy of the electrons.

For a complex interface that cannot be satisfactorily modeled by the dipolar Coulomb field, the angular dependence is either mapped out by electron ray tracing or determined *in situ* from the Coulomb refraction effect itself. For the latter, a low-order polynomial function is used to represent the angular dependence that is refined by correcting the anisotropic distortion of the Bragg diffraction patterns. introduced by interfacial Coulomb forces. Iterating the image symmetrization and Fourier phasing permits simultaneous determination of both surface potential and structural changes from the diffraction pattern.

**Ultrafast Diffraction**

Snapshots of diffraction images associated with transient structural changes induced by laser excitation are obtained following the pump-probe paradigm (Srinivasan et al., 2003). The temporal resolution of the diffraction images is defined by the effective pulsewidth of the electron beam – the 'shutter speed', determined based on the cross-correlation time between the pump and probe pulses, typically 200 fs to a few ps, depending on the incidence angle and beam sizes. A sequence of probe delays are programmed to have electron pulses arrive either before (negative time frame) or after (positive time frame) the pump laser pulse, at time intervals adjusted to match the rate of relative structural changes. The dynamics were recorded up to 3.5 ns to monitor the long-time diffusive relaxations in contrast to the short time photo-driven dynamics, and to ensure recovery at 1 ms interval by checking the reproducibility of the consecutive ground state images. For each time-resolved diffraction image, the accumulated image acquisition time is typically from 20 to 100 sec at 1 kHz repetition rate. The diffraction patterns were collected typically over scattering range $s$=1-15 Å$^{-1}$, from which we could extract *in-situ* measurement of the transient structure, temperature and charge of the nanoparticles in a manner to be discussed below.

**APPLICATIONS**

**Photoinduced nonhomogeneous structural transformation of gold nanocrystals**

The richness of diverse structures of gold and silver clusters and their experimentally observed interconversion, has prompted considerable efforts over two decades since the unexpected discovery of noncrystallographic structure types  (Marks, 1994; Mackay, 1962) and their transformations (shape fluctuations [Ijima & Ichihashi, 1986], and pre-melting [Ercolessi et al., 1991]). Whetten and coworkers (1996) researched gold clusters passivated by alkylthiolates and found the clusters to form discrete sizes with unusual stability. The possibility of directly observing shape changes of the structures was first demonstrated by high-resolution transmission electron microscopy (HRTEM) using electron beam irradiation as the stimulation source (Williams, 1987). Of particular interest in investigating such structural changes is the phenomena of 'surface premelting', where an early formation of a liquid layer over a solid core leads to the overall melting of the particle. This mechanism has been used to explain the shape transformation of Au nanorods (Link & El-Sayed, 2001), a premature drop in reflectivity (Plech et al., 2004), saturation of the acoustic frequency (Hartland et al., 2003), and coalescence of Au nanoparticles under pulsed laser excitation at a temperature below the estimated melting temperature (Plech et al., 2007). Whereas these observations are consistent with earlier theoretical studies (Ercolessi et al., 1991; Lewis et al., 1997), the inhomogeneous atomic dynamics of surface melting phenomena have never been directly recorded.

Here, we present a UEnC study of the structural dynamics of photoexcited 2nm Au nanoparticles (Ruan et al., 2007). The excitation wavelength is chosen at 800 nm to be far away from the surface plasmon resonance (SPR) bandwidth of the nanoparticles so as to avoid possible SPR-assisted photo-mechanical modification of the samples (Raman et al., submitted). Using diffraction difference method, we examine the changes following impulsive laser heating of the nanoparticles, in (a) reciprocal space, to assess the loss of long-range order based on Bragg peak analysis, and (b) in real space, to assess the change of local order based on the depletion and emergence of correlation density peaks in *G(r)*. The effective surface potential $V_s$ is also derived concurrently from the diffraction pattern using Fourier phasing method.

As shown in Fig. 12(a), the 'melting' occurs in three stages: (1) at 0-20 ps, the intensity of the Bragg peaks drops as a function of time, suggesting a loss of medium-range order associated with the 2nm Au nanocrystals. (2) from 20 to 200 ps, the depletion of the Bragg peak appears to be split, with shoulders emerging as less negative and gradually becoming positive peaks, suggesting a coexistence of disordered domains and partial recovery of medium-range orders. (3) from 335 ps onwards, the negative peaks largely disappear, with some weak positive peaks standing out in positions similar to the ground state Bragg spectrum, signifying the existence of an enhanced transient order at an elevated temperature.

The loss of medium-range order in the 0-20ps stage could be explained in one of two ways - a thermal disorder or a reduction in the persistence length due to partial melting. Thermal disorder in crystals is often treated within the frame-work of Debye-Waller analysis, which is based on the concept of random, homogeneous variations of bond distances throughout the crystal. In this, the drop of the Bragg diffraction intensity is associated with $\bar{u}_\perp^2$, the mean square displacement (MSD) of atoms perpendicular to the corresponding Bragg planes through the relation $\ln\left(I_s(t)/I_s(t<0)\right) = -s^2 \Delta \bar{u}_\perp^2(t)/4$. Assuming the displacements of atoms are perfectly isotropic around the equilibrium positions, then the MSD=$3\bar{u}_\perp^2$. Thus, one can use the drop of diffraction intensity to measure MSD, and hence the temperature of the crystal. We perform the Time-Dependent Debye-Waller (TDDW) analysis on the groups of (220), (311) and (331) Bragg peaks, and find that an anisotropy exists between them, as shown in Fig.13. Different Bragg peaks yield a different value for the MSD which cannot be explained by a homogeneous thermal disorder model. Further, the fact that high-order diffraction peak (331) exhibits more disorder than the lower order ones suggests that the longer atom-atom correlations are more perturbed than the shorter ones, signifying the reduction of the persistence length in nanocrystals.

To examine the local structures, we turn our attention to the *G(r,t)*. In comparison with the ground state *G(r*, -1 ps), the difference curves ΔG(r, t) in Fig. 12(b) reveal that from 10-20 ps, the correlation density above 10 Å is almost wiped out completely, while correlation densities corresponding to the nearest neighbor bonds (2.89-7.50 Å) deplete proportionally much less. This is a clear indication of the lattice disorder with significantly reduced persistence length. By measuring the expansion of nearest neighbor distances, we can tentatively assign a lattice temperature based on the thermal expansion coefficient (Touloukian et al., 1975). We find that the maximum lattice temperature in Fig. 13 approaches 950 K, which is close to the steady-state melting temperature of ~900 K (Buffat & Borel, 1976) for 2 nm Au nanocrystal, but above the reported 'surface melting' temperature of 377 K (Plech et al., 2007). We also examine the coevolution of *Vs* and find it to proceed on a different timescale from the thermal expansion. Thus by combining insights from *S(s)* and *G(r)* analysis, we can tentatively attribute the loss of the long-range order and the reduction of the persistence length to a 'nonhomogeneous premelting' phenomenon. However, without quantitative 3D modeling, we cannot differentiate 'phase coexistence', where the nanoparticle breaks into homogeneously distributed crystalline and disordered regions, and 'surface premelting', where a continuous liquid layer forming at the surface, which thickens as temperature increases (Wang et al., 2008; Ercolessi et al., 1991). Net growth of correlation densities at slightly longer distances relative to those of the ground state structure is clearly visible on the nanosecond time scale.

**Four-dimensional structural refinement using Progressive Reverse Monte Carlo method**

The as-derived pair-correlation function $G(r)$ is a one-dimensional (1D) projection of the three-dimensional atomic structures, containing only the first-order atom-atom correlation. To retrieve higher order atom-atom correlations, hence recovering the 3D atomic structure, a refinement of the experimental curves with simulations based on an atomic structural model is required. For nanostructures, the 3D refinement usually contains more structural parameters than the constraints from the powder data. To avoid ambiguity in the structural determination based on a down-hill search, a "complex modeling" paradigm that combines theory and experiment in a self-consistent computational framework is used to solve the nanostructure problems (Billinge & Levin, 2007). Fortunately here, for a small homonuclear Au nanocrystal, its ground-state refinement is well constrained by the lattice symmetry to within a resolution of 0.05 Å. Nevertheless, for excited-state structure with significant disorder, the search will indeed diverge as the number of possible solutions increases and a proper constraint to restrict the refinement space is necessary. A solution to this 3D structural refinement problem is based on a Progressive Reverse Monte Carlo (PRMC) algorithm, which uses simulated annealing for structural refinement with a perturbative interation scheme. Known to be a global structural solver, but prone to yielding non-unique solutions, the Reverse Monte Carlo (RMC) algorithm relies on physical constraints to guide the search (McGreevy, 2001). We have implemented a scheme to restrain the RMC search for solving the UEnC structural problem, based on the diffraction referencing in UEnC. This is done by choosing the time interval between neighboring frames in the UEnC data to match the rate of change such that the structural difference between adjacent time frames is small. This maintains the structural correlation between the frames, thereby allowing changes to be tracked progressively and reliably within a restrictive search space, without need for a global search, as depicted in Fig. 14. Using this approach, the 3D representation of the transient atomic structures of gold nanocrystals is constructed. We start with a supercell model of a 2.5 nm cuboctahedral nanocrystal with a cubic cell size of 50 Å. The PRMC algorithm iteratively fit both the simulated $G(r)$ and s$M(s)$ with experimental counterparts based on adjusting the supercell structure. The cost function is based on the sum of $\chi^2$ values fitting the normalized $G(r)$ and $sM(s)$ (with equal weight). The results are depicted in Fig. 15 for selected time frames. These 3D snap-shot representations of the excited structures display features that are important to differentiate different melting scenarios. The appearance of an evolving under-coordinated surface layer can clearly be identified. As the temperature rises, the under-coordinated layer thickens, causing a contraction of the crystalline core. At the peak of the surface melting at ~ 15-20 ps, less than 25% of the atoms remaining structured within the core region. Using the core size as an index, we can trace the propagation of melting and re-solidification fronts. This 3D refinement scheme not only visualizes the assessment based on the 1D diffraction difference analyses made earlier, but also opens up the prospects of extracting other higher-order correlations within the atomic framework to study each step along the reorganization of the local bonding network during phase transformations.

**Graphite-to-diamond transformation**

One particular interest in nanocarbon material research is the conversion of graphite to diamond, which is believed to involve the rhombohedral phase of graphite as an intermediate state (Fahy et al., 1986; Yang & Wang, 2001). While these two forms of carbon are nearly degenerate energetically, they are separated by a formidable energy barrier (Nakayama & Katayama-Yoshida, 2003; Fahy et al., 1986). Only under environments of extreme heat and pressure can the interconversion occur thermodynamically. Recent theoretical studies have suggested lowering of this energy barrier by charge doping (Nakayama & Katayama-Yoshida, 2003) or electronic excitation (Ishioka, et al., 2001). This has been supported by the experimental discovery of nano-diamond formation at ambient temperatures by impinging a highly charged Argon ion beam ($Ar^{8+}$) on graphite, followed by charge injection from an STM tip (Merugo et al., 2001).

To explore a plausible all-optical pathway for the same, we studied photoinduced lattice motion in HOPG using ultrafast electron crystallography (Raman et al., 2008). HOPG is made of two-dimensional graphite micro-domains with randomly oriented basal planes, which we excite at 800nm. This excitation is strongly coupled to the interlayer shearing $E_{2g}$ phonon mode (Mishina et al., 2000) and can induce a change of layering symmetry of graphite from the hexongal (ABAB) towards the rhombohedral (ABCABC) phase required for the formation of interlayer $sp^3$ bonds (Scandolo et al., 1995). The ground state UEC pattern from HOPG consists of (0,0,6), (0,0,8), (0,0,10) and (0,0,12) interlayer Bragg interferences. Performing Time-Dependent Debye-Waller analysis on these, we find the timescale for the mean-square displacement to be 8±1 ps, which is consistent with the thermalization timescale of the initial non-equilibrium $E_{2g}$ phonon modes with the rest of the phonon bath (Kampfrath et al., 2005). We also observe a reduction of the (0,0,6) and (0,0,10) Bragg peaks relative to (0,0,8) and (0,0,12), as shown in Fig. 16, corresponding to a halving of lattice constant. This is accompanied by emergence of a well-defined interlayer bond at $r = 1.9$Å in the layer density function LDF of graphite, which corresponds to a new interlayer spacing. The disappearance of this peak at long times points towards the creation of a reversible $sp^2$-$sp^3$ hybrid structure, similar to that identified in another pressure driven graphite-to-diamond transformation in a diamond anvil cell (Mao et al., 2003). This diamondization process is likely driven by a compressive Coulomb stress created by the photoinduced charge separation near the surface induced through hot electron injection into surface image potential states (Fig. 4).

**Interfacial charge dynamics and molecular nanoelectronics**

The capability of time-resolved diffraction to determine photovoltages has been demonstrated recently for an Si/SiO$_2$ interface (Murdick et al., 2008). A ~2nm SiO$_2$ layer is chemically grown on a clean silicon surface. This interface has previously been investigated using time-resolved photoemission methods (Marsi et al., 2000), showing a transient photovoltage that is generated by the tunneling of photoexcited carriers from silicon across the SiO$_2$ layer to the surface states. The surface charging causes a new barrier potential to be established within the silicon underlayer by rearranging the space-charge according to the surface state population. The electric field established within the barrier region and the insulating SiO$_2$ barrier cause the coherently diffracted electron to be deflected. Using the dipolar field model as described by Eqn. (6), we can determine the photovoltage as a function of time. Since the chosen laser excitation energy is near infrared (800nm), the optical penetration depth is several microns in the Si substrate. In turn, the Bragg peak movements due to thermal expansion are negligible compared to those arising from Coulomb refraction. We have shown that the photovoltage dynamics are consistent with surface state relaxation investigated by time-resolved photoemission (Halas & Bokor, 1989), suggesting that surface charge recombination is driven by drift-diffusion in the space-charge region, which can be modeled by a power-law decay with exponent –1 (Murdick et al., 2008).

The relatively long time (100 ps), but rechargable, surface charges are ideal for an array of studies involving charges at interfaces. One subject of particular interest to nanoelectronic and catalysis research is the mechanism of quantum tunneling into a quantum dot or transport through a molecular wire, down to single charge transfer level. Investigation is typically done under steady-state conditions using fabricated nanointerfaces (Sato et al., 1997b) or a scanning probe (Wang et al., 2005). By gating the diffraction at regions where it is dominated by molecular interferences pattern from the SAM, we are able to determine the molecular structure and its transient response to the imposing interfacial potential change (Fig. 17), determined by the charging ($q$) of the nanoparticle. The clear deviation of the transient Coulomb refraction response from the molecular diffraction relative to that of the silicon diffraction is clear evidence of electrons advancing from the Si interface states into the nanoparticle through the molecular wires. Thus the observed changes are indicative of charging and discharging of the nanoparticles. Based on the Coulomb refraction shift for 20 nm Au nanoparticle, which is on the order of 20V/ $\text{Å}^{-1}$, and the electron ray tracing across the gap region between the nanoparticle and the silicon substrate, we estimate that the charging is 250 $e^-$ per $\text{Å}^{-1}$. With the sensitivity of $3\times10^{-3}$ $\text{Å}^{-1}$ we have tracking the diffraction shift we should be able to probe single charge transfer events in small nanoparticles within the Coulomb blockade regime.

**CONCLUSION**

The development of an ultrafast electron crystallography method for quantitative structural determinations of nanostructures and interfaces are outlined. Its sensitivity and resolution are demonstrated in the studies of surface melting of gold nanocrystals, non-equilibrium structural transformation of graphite, and molecular responses to interfacial charge transfer, showing a versatility for not only determining the structures, but also the charge and energy redistribution at interfaces. The ability of efficient probing of widely dispersed nanostructures, as low as 6 particles/$\mu m^2$, using a focused probe, demonstrates the possibility of single particle detection using UEnC. With the emerging sub-micron sized probe, a new territory that combines imaging, diffraction, and spectroscopy is within reach using the UEnC platform for investigating surface-supported nanostructure dynamics.

**ACKNOWLEDGMENTS**

We would like to express our gratitude to Prof. A.H. Zewail and Prof. M. Fink for inspiration of this work. This work is an extension of the UEC development made first in AHZ's laboratory at Caltech when CYR was a postdoctoral scholar. We also acknowledge Prof. S.J.L. Billinge, Prof. P.M. Duxbury, for sharing their insights on nanostructure analysis, Mr. M.A. Khasawneh for his contribution in the early phase of the work, and MSU Physics machine shop for fabricating the instruments. This work is supported by US Department of Energy under Grant DE-FG02-06ER46309, National Science Foundation under NSF-DMR Grant 0703940, American Chemical Society's Petroleum Research Fund under Grant 45982-G10, and MSU Intramural Research Grants Program.

To appear in the special issue of 'Microscopy & Microanalysis' on ultrafast electron microscopy and ultrafast sciences (Vol 15, 323-337, 2009).DWYER, J.R., HEBEISEN1, C.T., ERNSTORFER1, R., HARB, M., DEYIRMENJIAN, V.B., JORDAN, R.E. & MILLER, R.J.D. (2006). Femtosecond electron diffraction: 'making the molecular movie'. *Phil Trans R Soc A* **364**, 741-778.

EBERHARDT, W. (2002). Clusters as new materials. *Surf Sci* **500**, 242-270.

ERCOLESSI, F., ANDREONI, W. & TOSATTI, E. (1991). Melting of small gold particles: Mechanism and size effects. *Phys Rev Lett* **66**, 911 – 914.

FAHY S., LOUIE S.G. & COHEN, M.L. (1986). Pseudopotential total-energy study of the transition from rhombohedral graphite to diamond. *Phys Rev B* **34**, 1191-1199.

HAKKINEN, H., ABBET, W., SANCHEZ, A., HEIZ, U. & LANDMAN, U. (2003). Structural, electronic, and impurity-doping effects in nanoscale chemistry: Supported gold nanoclusters. *Angew Chem Int Ed* **42**, 1297-1300.

HARGITTAI I. AND M. HARGITTAI (1988). *Stereochemical applications of gas-phase electron diffraction.* (Wiley-VCH, New York).

HALAS, N.J. & BOKOR, J. (1989). Surface recombination on the Si(111) 2x1 surface. *Phys Rev Lett* **62**, 1679-1689.

HARTLAND, G. V., HU, M. & SADER, J. E. (2003). Softening of the symmetric breathing mode in gold particles by laser-induced heating. *Phys Chem B* **107**, 7472 – 7478.

HARUTA, M. (1997). Size- and support-dependency in the catalysis of gold. *Catalysis Today* **36**, 153-166.

HOMMELHOFF, P., SORTAIS, Y., AGHAJANI, A.-T. & KASEVICH, M.A. (2006). Field emission tip as a nanometer source of free electron femtosecond pulses. *Phys Rev Lett* **96**, 077401.

IIJIMA, S. & ICHIHASHI, T. (1986). Structural instability of ultrafine particles of metals. *Phys Rev Lett* **56**, 616 - 619.

INO, S., & OGAWA, J. (1967). Multiply twinned particles at earlier stages of gold film formation on alkalihalide crystals. *J Phys Soc Jpn* **22**, 1365-1374.

ISHIOKA, K., HASE, M. KITAJIMA, M. & USHIDA, K. (2001). Ultrafast carrier and phonon dynamics in ion-irradiated graphite. *Appl Phys Lett* **78**, 3965-3967.

KAMPFRATH, T., PERFETTI, L., SCHAPPER, F., FRISCHKORN, C. & WOLF, M. (2005). Strongly coupled optical phonons in the ultrafast dynamics of the electronic energy and current relaxation in graphite. *Phys Rev Lett* **95**, 187403.
18

**FIGURES AND CAPTIONS**

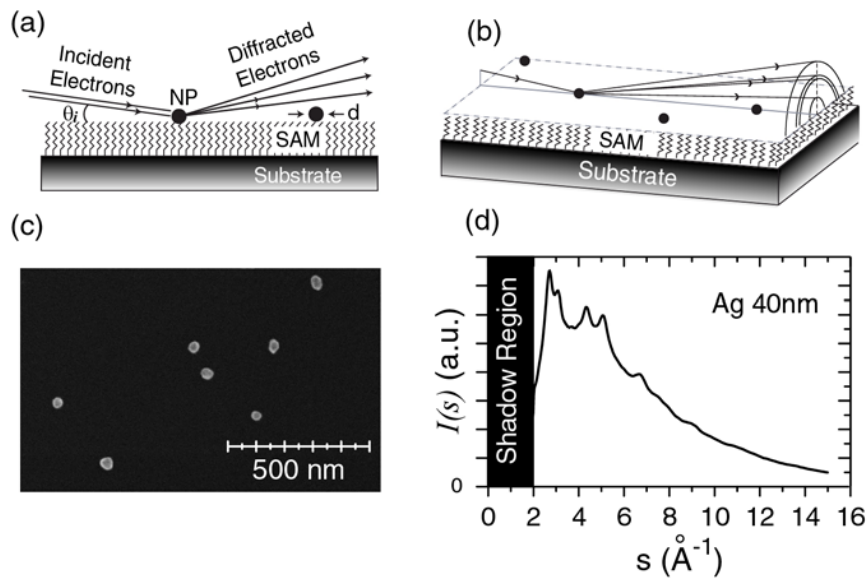

**Fig. 1**. Nanopowder diffraction arrangement for surface supported nanoparticles. (a) Scanning electron microscopy image of Ag nanoparticles (d=40 nm) dispersed on a Si(111) surface. (b,c) At a grazing incident angle $\theta_i \approx 1\text{-}5°$, the electron beam is used to diffract from the nanoparticles supported on a surface with a scattering solid angle up to $2\pi$. To minimize the interference from the substrate scattering, a 'soft' buffer layer is employed to elevate the nanoparticle from the substrate while suppressing the background scattering. (d) Total diffraction curve from such a diffraction geometry.

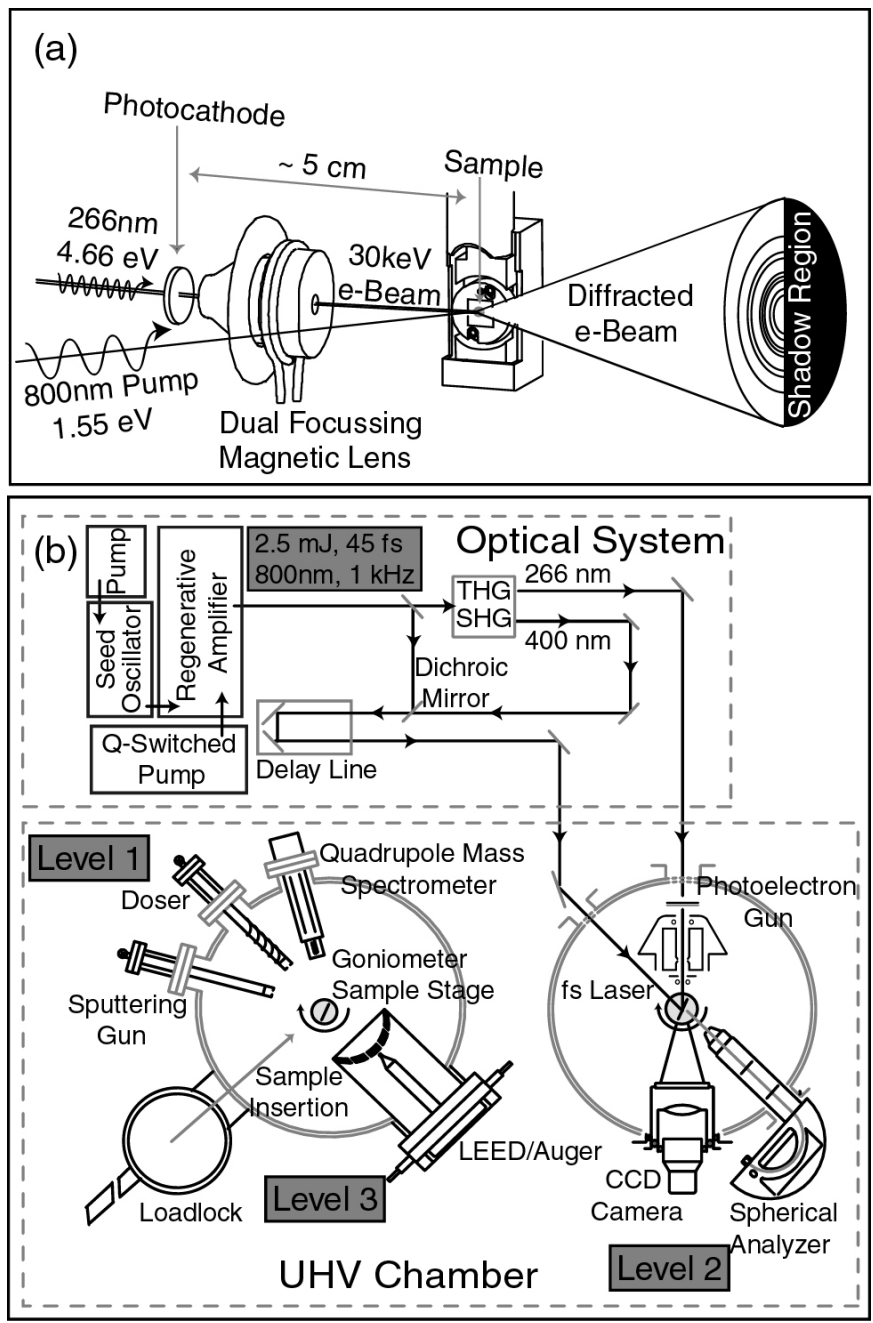

**Fig 2.** The UEnC Pump-Probe Apparatus. (a) Experimental setup showing the laser pump – electron probe geometry. (b) Layout of the UHV chamber along with the femtosecond optical setup.

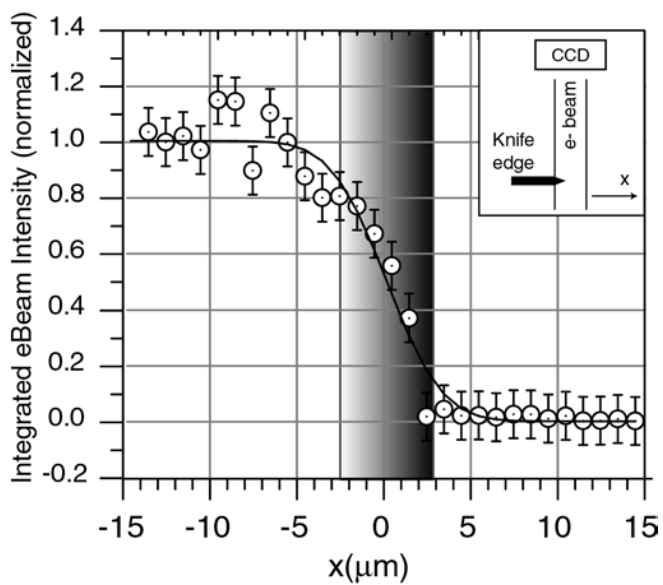

**Fig 3.** Measurement of electron beam size at the sample location using the 'knife-edge' method (inset).

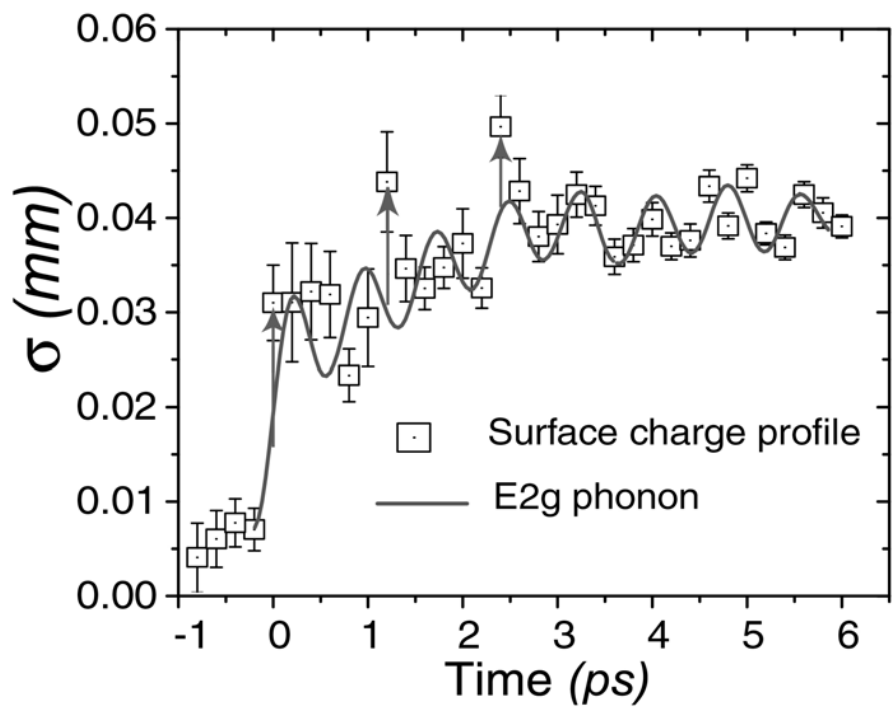

**Fig 4.** Detection of fs jump in the surface charge profile in HOPG following photoexcitation, and its subsequent oscillation, with a sinusoidal fit corresponding to the 0.78 ps period of the $E_{2g}$ phonon mode, highlighting the sub-ps instrumental time resolution.

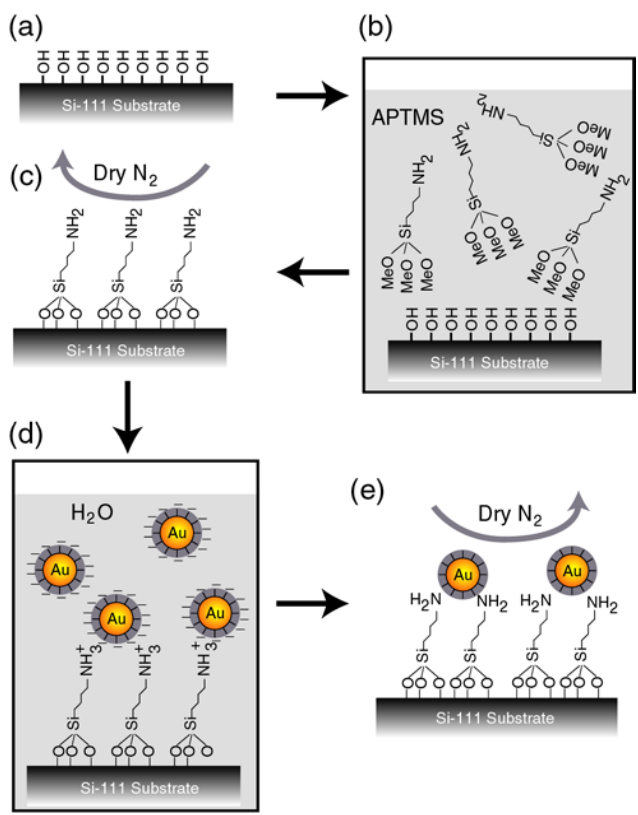

**Fig 5.** Schematic representation of steps used in preparing the nanoparticle samples for UEnC experiments. These steps involve (a) cleaning the Silicon substrate using a modified-RCA method, (b), (c) surface functionalization and (d), (e) nanoparticle dispersion (see text).

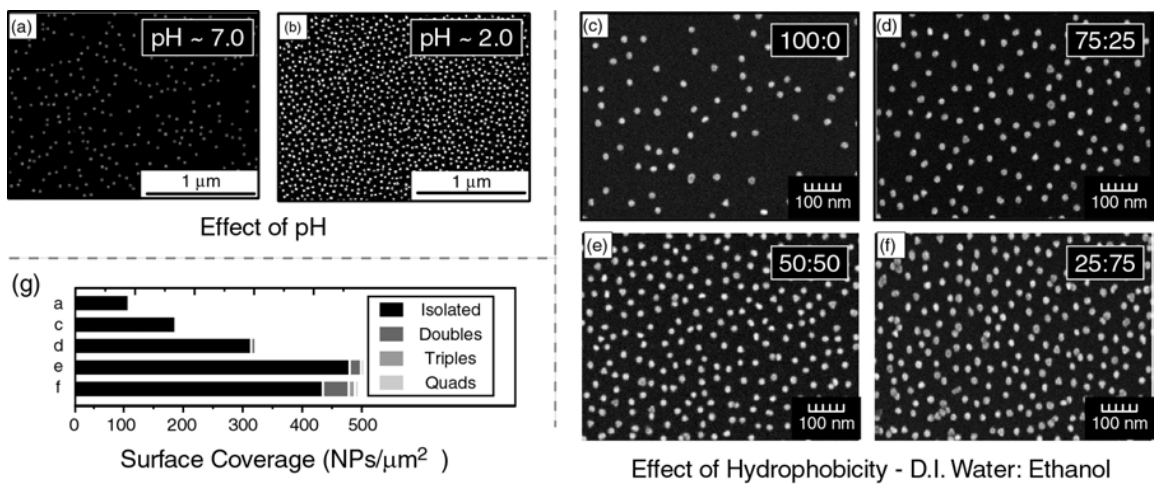

**Fig 6.** SEM micrographs showing controlled deposition of gold nanoparticles on Si-111 surface, achieved by controlling the pH of the gold colloid and hydrophobicity of the functionalised Si-111 surface. The bar-graph indicates the increase in particle agglomeration at increased surface coverage.

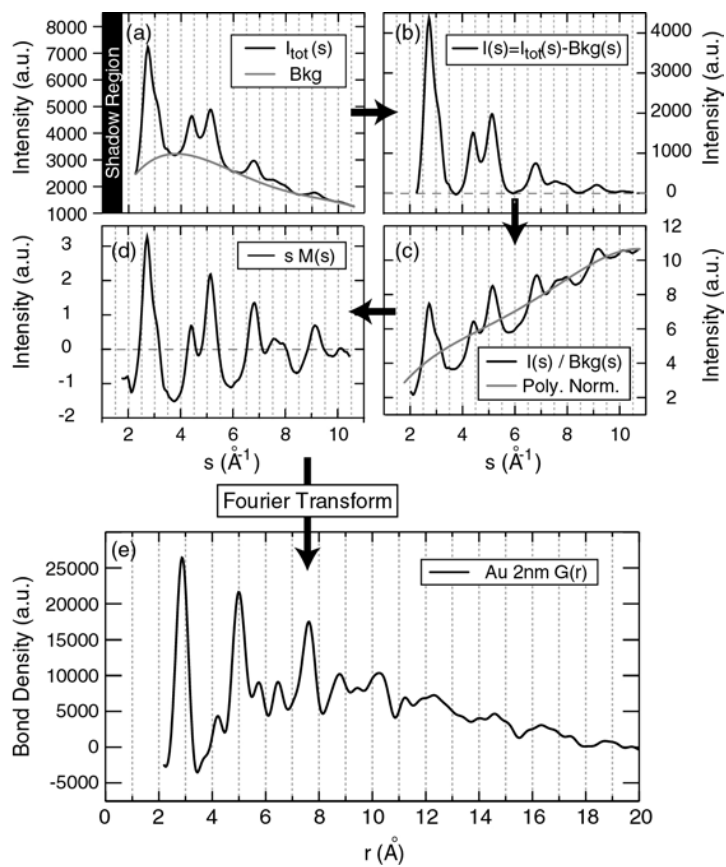

**Fig 7.** UEnC Data reduction procedure highlighting the sequential steps to obtain the structure function *S(s)*, pair distribution function *G(r)* starting from the total scattering intensity *I(s)*.

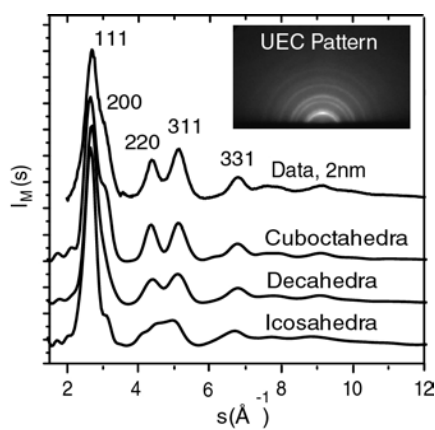

**Fig 8.** Experimentally obtained structure function *S(s)* for 2nm gold nanoparticles, compared to simulated curves from the known non-crystallographic motifs prevalent in nanostructures. The resemblance of experimental curve to a cuboctahedra structure is apparent.

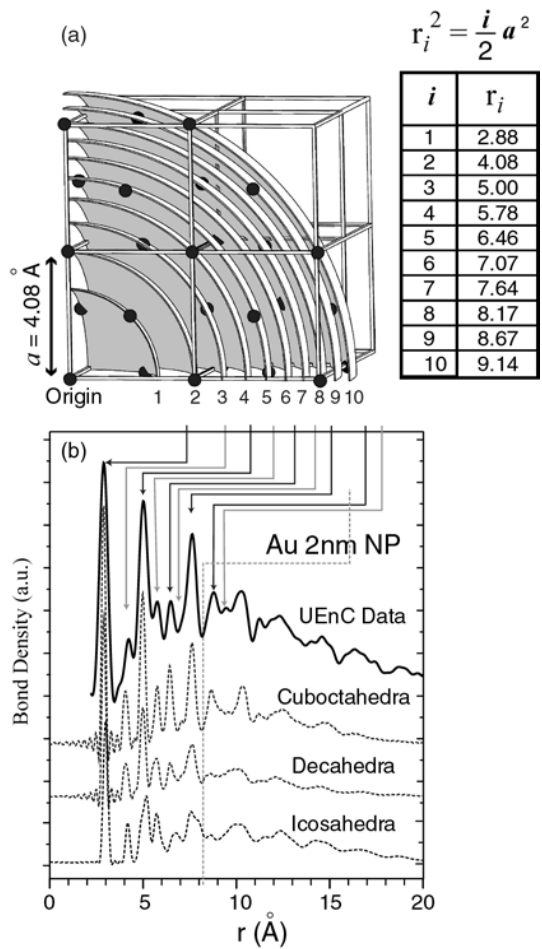

**Fig 9.** Structure analyses of 2 nm Au nanoparticles. (a) fcc coordination shells corresponding to interatomic distances $r_i$, calculated based on the bond order $i$ and the Au lattice constant, $a =$ 4.08 Å. (b) Experimental modified radial distribution functions of static Au NPs along with theoretical predictions for cuboctahedra, decahedra, and icosahedra.

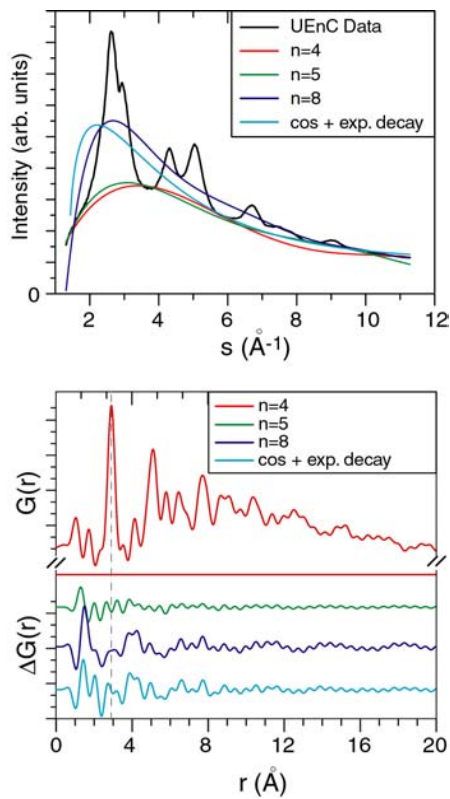

**Fig. 10.** Robustness of data reduction procedure. (a) Different background fits to the total scattering intensity I(s) obtained from UEnC. n indicates the order of polynomial fit. (b) G(r) for n=4 background fit, along with the residual G(r) for all cases, relative to n=4 case. The residuals are almost null in all cases, showing only minor noise fluctuations, thus proving that the different background fitting routines all yield the same G(r). The only region where the residuals are non-zero is at small r, where it involves contribution from diffusive molecular scattering. These low-density peaks are incommensurate with atom-atom correlation lengths of fcc structures, and hence can be rejected from refinement.

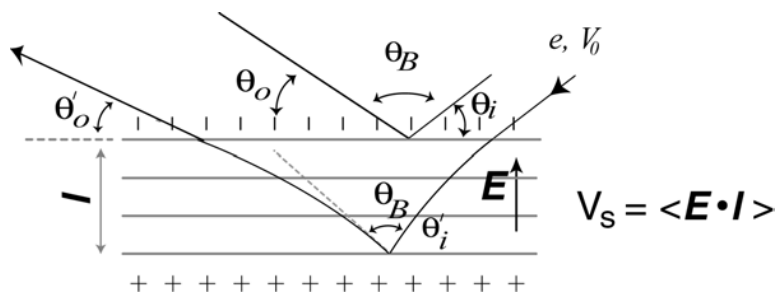

**Fig. 11.** Coulomb refraction effect on the diffraction pattern due to presence of photofield. $\theta_i$ is the incidence angle for the electrons, $(\pi-\theta_B)$ the Bragg angle at which the electrons scatter, $\theta_o$ the exit angle in absence of photofield E. $\theta'_i$ is the electron incidence angle, and $\theta'_o$ the electron exit angle in presence of **E**. $l$ denotes the probe depth of the electrons.

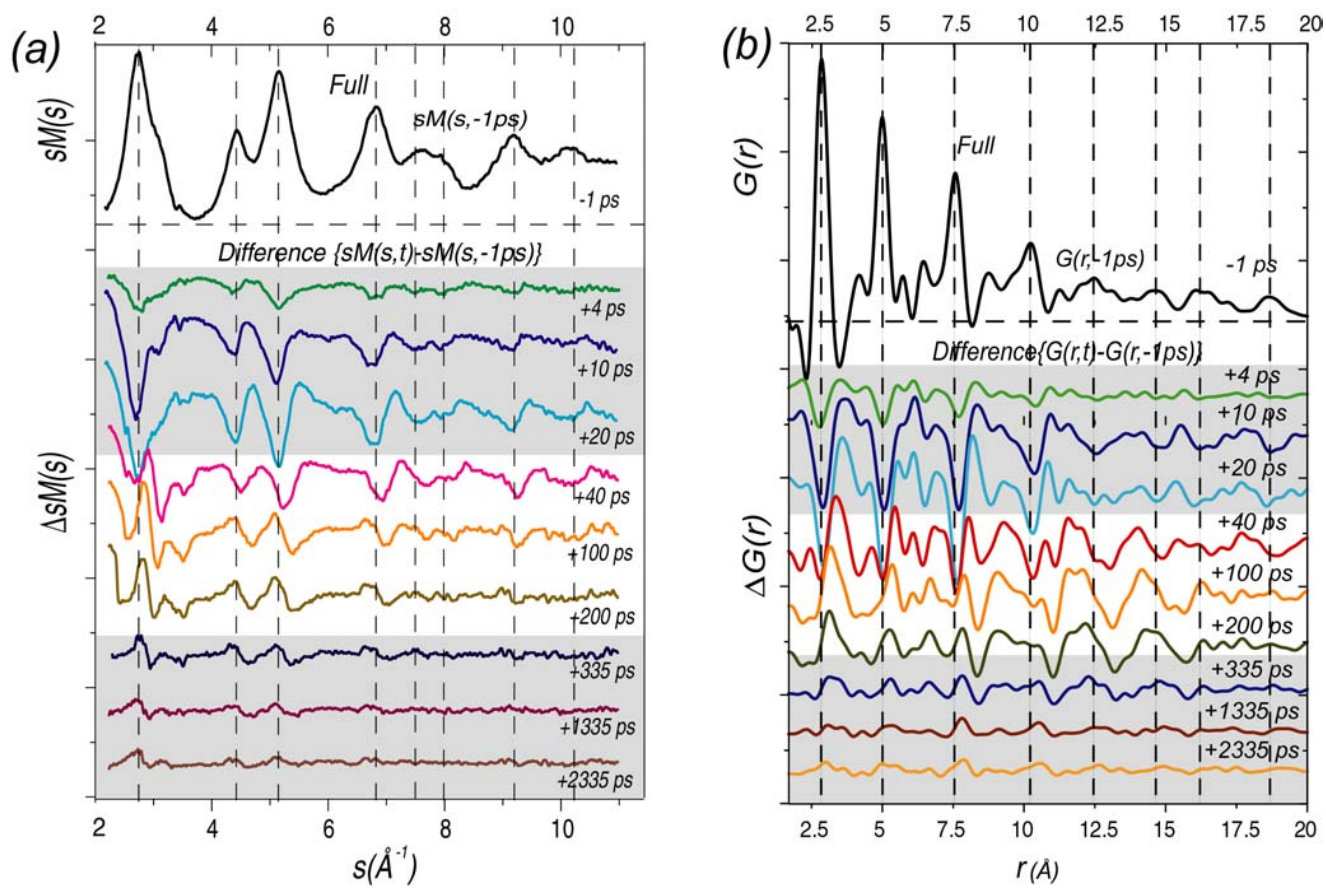

**Fig. 12.** Transient long-range (a) and local order (b) pictures in the melting of a 2 nm Au nanoparticle. (a) $sM(s)$ difference curves, showing loss of long-range order, evident from a drop in Bragg peak intensity (0 – 20 ps), followed by the coexistence of disordered domains with partial long-range order (20 – 200 ps). From 335 ps onward, the existence of an enhanced transient order at an elevated temperature is observed (see text). (b) $G(r)$ difference curves showing the lattice dis-ordering with a significantly reduced persistence length (see text).

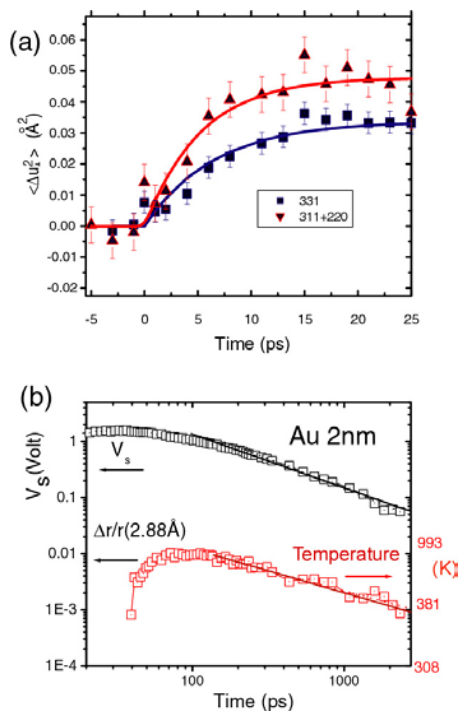

**Fig. 13.** (a) Projected mean square displacement of atoms ($\Delta \bar{u}_\perp^2$) measured from the drop in Bragg peak intensity through Debye-Waller analysis. An anisotropy between the peaks is clear, as peak (331) exhibits more disorder than the lower orders, signifying a reduced persistence length in nanocrystals (see text). (b) The transient surface photovoltage measured in the nanoparticle (left axis). Lattice temperature is deduced from the local bond stretches in conjunction with the thermal expansion coefficient (right axis).

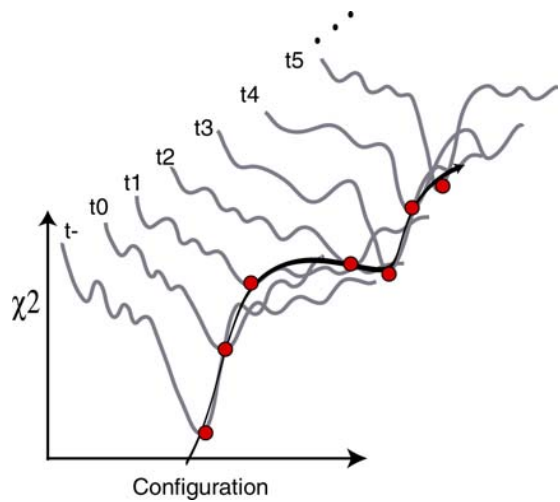

**Fig. 14.** Progressive Reverse Monte-Carlo Refinement (PRMC). By choosing suitable time-instances t0, t1, t2, … at which to probe the non-equilibrium structures, we ensure that the relative structural change from one to the next is not too drastic, allowing PRMC to reliably track the structural evolution.

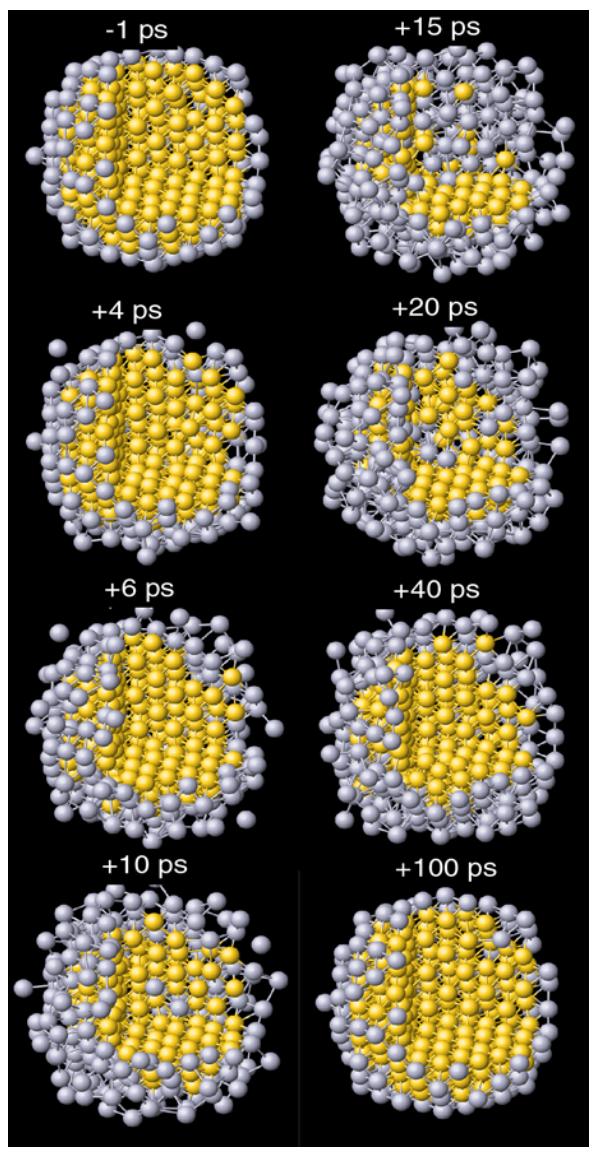

**Fig.15.** Reverse Monte Carlo (RMC) results. The atom-coordination at various delays suggests that melting commences through the route of "surface melting" rather than "partial melting."

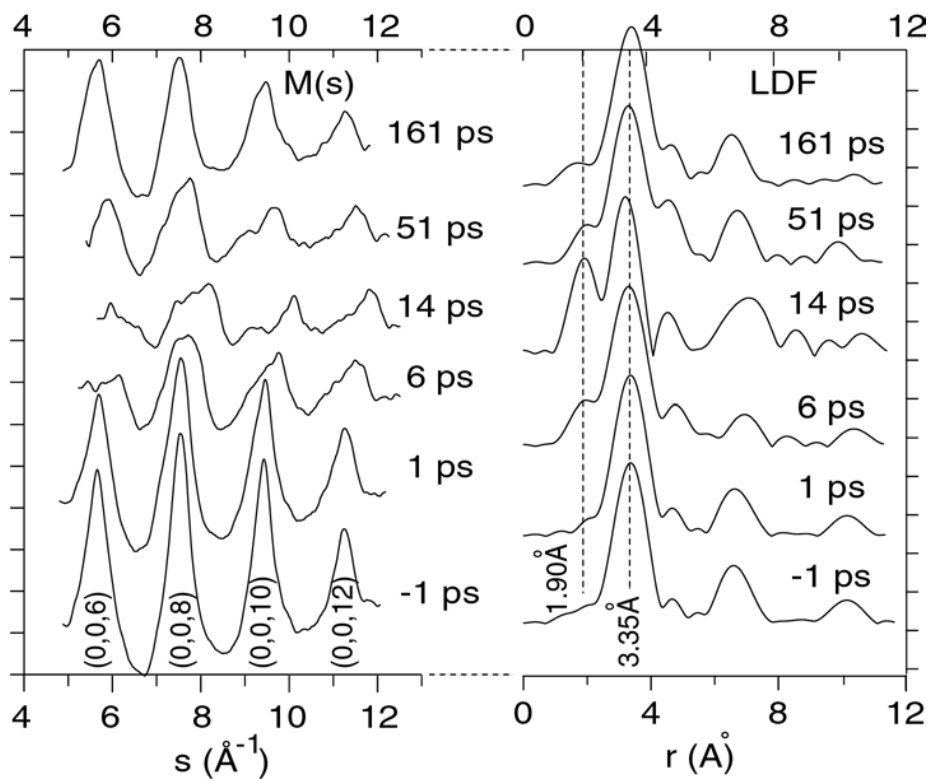

**Fig 16.** Molecular interference pattern *M*(*s*) and the corresponding Layer Density Distribution (LDF) curves at selected time instances following strong photoexcitation. The transient peak at 1.9 Å in the LDF indicates formation of an interlayer bond.

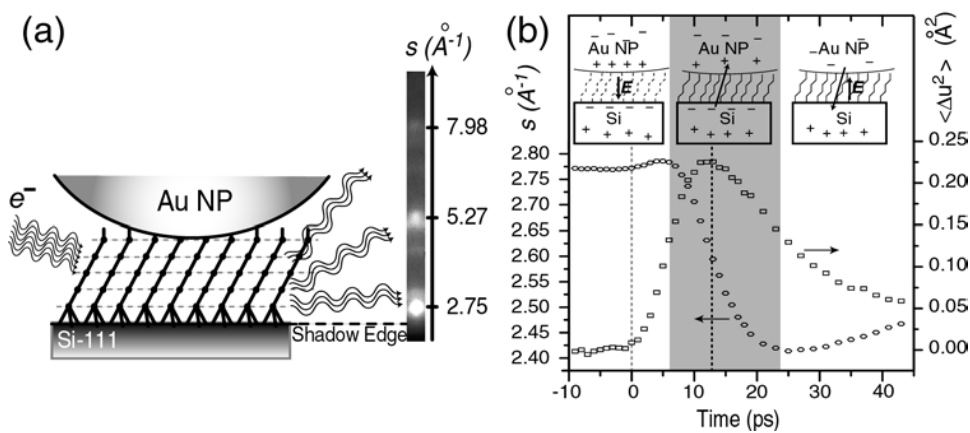

**Fig 17.** Electron diffraction from a molecular wire. (a) Schematic representation of the static structure of the SAM as the probe electrons are scattered from it. (b) Electron transport through the molecular wire suggesting a charging and discharging of the nanoparticle (see text).